# FUZZY INFORMER HOMED ROUTING PROTOCOL FOR WIRELESS SENSOR NETWORK

Gholamreza Kakamanshadi, Savita Gupta and Sukhwinder Singh

Department of Computer Science & Engineering, Faculty of Engineering and Technology, Panjab University, Chandigarh, India

## ABSTRACT

*A wireless sensor network consists of severalsensor nodes. Sensor nodes collaborate to collect meaningful environmental information and send them to the base station. During these processes, nodes are prone to failure, due to the energy depletion, hardware or software failure, etc. Therefore, fault tolerance and energy efficiency are two important objectives for reliable packet delivery. To address these objectives a novel method called fuzzy informer homed routing protocol is introduced. The proposed method tries to distribute the workload between every sensor node. A fuzzy logic approach is used to handle uncertainties in cluster head communication range estimation. The simulation results show that the proposed method can significantly reduce energy consumption as compared with IHR and DHR protocols. Furthermore, results revealed that it performs better than IHR and DHR protocols in terms of first node dead and half of the nodes alive, throughput and total remaining energy. It is concluded that the proposed protocol is a stable and energy efficient fault tolerance algorithm for wireless sensor networks.*

## 1. INTRODUCTION

A wireless sensor network (WSN) is a self-configured or self-organized network. It contains a collection of small, low powered sensor nodes with limited transmission range and the base station or sink [1, 2]. Sensor nodes in wireless sensor networks are prone to failure because of hardware and software failures, instability of communication link, battery depletion, dislocation, etc. Therefore, there is a needfor an efficient fault tolerance mechanism to manage or identify the fault and take appropriate action while it occurs [3]. A collector node has restricted capability in sensing and collecting meaningful environment information within its range. It generally transfers the sensed and collected data to the base station. The sensor nodes consume energy while sensing, processing, receiving and transferring data [4]. In the majority of cases, they have the same amount of energy which is not replaceable [5] or replacing the battery is impossible [6, 7]. Hence, energy efficiency is an important design objective in a wireless sensor network.

To aggregate and transmit sensed data through efficient manner, the network can be clustered or partitioned into the number of clusters. Each cluster in the network has a cluster head. It generally receives the sensed data from cluster members then aggregates and transmits to the base station [8].In the clustering manner, selection of a suitable cluster head is very important, it can reduce energy depletion of sensor nodes and increase the lifetime of the wireless sensor network [9, 13]. Generally, clustering procedures utilize two methods, electing cluster heads with higher remaining energy and revolving them periodically to balance energy depletion of the sensor nodes all over the network for prolonging the network lifetime.

Utilizing intelligent methods improve the efficiency of wireless sensor network [10]. As an intelligent technique, fuzzy logic is the most powerful tool that can be used for clustering procedure. In a wireless sensor network, it can be used to select suitable cluster heads [9, 11]. Fuzzy logic has several advantages over traditional methods for instant, design time, computational complexity and development cost are low and it is more flexible [12].Sointhis paper, a novel fault tolerance algorithm named Fuzzy Informer Homed Routing (FIHR) protocol is introduced and simulated for wireless sensor networks using fuzzy logic. It is derived from the Informer Homed Routing (IHR) protocol [13] and Dual-Homed fault tolerant Routing (DHR) protocol [14]. Clustering procedure in IHR and DHR protocols is based on a probabilistic model same as LEACH protocol. Moreover, it is probable that some cluster heads are located in a particular zone. It means that primary cluster heads are not picked out in a distributed manner. The proposed FIHR is a distributed competitive cluster head selection with fault tolerance algorithm. The proposed method efforts on allocating suitable communication range to the tentative primary cluster heads. To make wise decisions to select primary cluster heads (PCHs), the introduced FIHR method employs fuzzy logic inference system and uses the distance to the base station and the remaining energy of available sensor nodes during simulation time. Furthermore, each primary cluster head will choose the non-cluster head locally with the higher energy left as its backup cluster head (BCH). To achieve fault tolerance, every BCH will control the aliveness of relevant PCH based on the beacon message it receives from its PCH in each round.

We compare the effectiveness of FIHR protocol with IHR and DHR protocols in terms of first node dead, half of the nodes alive and total residual energy level of the network at various rounds. The results show that the proposed FIHR protocol outperforms IHR and DHR protocols. It is concluded that the suggested FIHR protocol is stable as well as energy efficient fault tolerance protocol for WSNs.

The rest of this article is organized as follows: Section 2 is about related works, in section 3 Preliminaries will be discussed, in section 4 the proposed FIHR protocol will be introduced, in section 5 clustering with fuzzy logic system will be discussed, section 6 is about simulation results and discussions, and finally we conclude the paper in section 7.

## 2. RELATED WORK

Efficiently transferring the data from collector nodes to the sink or base station is a critical issue in the wireless sensor network. Therefore, numerous faulttolerant routing protocolshave been offered in the literature. Moreover, most of the existing fault tolerance methods introduce hardware redundancy and path redundancy. For example, to offer fault tolerance against cluster head failures, DHR protocol was proposed. In this method, each cluster is structured into collector nodes and two cluster head nodes (primary and backup). Furthermore, cluster head selection procedure is performed in rounds as in Low-Energy Adaptive Clustering Hierarchy (LEACH) protocol. In this protocol the collected data is dispatched to both cluster heads then the primary and backup cluster heads send received data packets to the base station [14] subsequentlyinsignificant energy will be consumed per packet transfer. Moreover, a major drawback of DHR protocol is a duplication of forwarding every data packet over two disjoint paths towards thebase station. This cause decreases the overall network lifetime. Furthermore, the DHR protocol uses a simple probabilistic model which is insufficient to find the best solution for network clustering. Qiu et al. [13] presented a novel energy-aware and fault tolerance scheme for a wireless sensor network, calledInformer Homed Routing protocol. In this protocol, each collector node has two cluster heads associated with it (PCH and BCH). Furthermore, the collector node just dispatched sensed data to the BCH when it found that PCH failed, instead of transferring the data packet to the PCH and BCH simultaneity. The effectiveness of this method was compared with DHR and LEACH protocols in aspects of power consumption and the number of failed sensor nodes. Also, the throughput of the network was considered with different fault rates. Results revealed that the suggested protocol could significantly decrease power consumption and reduce data packet loss rate as well as prolong the networklifetime. IHR protocol does not measure the remaining energy level of the selected cluster head during cluster formation. Moreover, the suggested protocol uses a simple probabilistic model which is not enough to gain the best solution for clustering. Furthermore, the IHR protocol does not take into considerations the distance among primary cluster heads.

Abedi et al. [15] introduced a new fault tolerance algorithm. In this algorithm, to guarantee a fault tolerant topology, each sensor node must select two nearest relay nodes and can be considered as primary and backup cluster heads. Furthermore, the failure probability value of each relay node was used to discover the most optimal path in the current network topology. Moreover, the relay node with the lowest probability of failure will become the primary cluster head and the second one is considered as a backup. Therefore, sensor nodes could send a data packet through the suitable relay nodes to the sink and the event of primary cluster head failure, sensor nodes could send a data packet through the backup cluster head. The simulation results were used to examine only the fault tolerance level of the networks. There is a limitation of this method, sensor nodes select the next hop relay node without taking into account the total distance among itself and the base station also does not consider the residual energy of them.

## 3. PRELIMINARIES

To describe the suggested protocol in detail, the characteristics of the network model that are used in the simulation process are introduced. Thus, the following assumptions that are made about the wireless sensor network model are given below:

- Sensor nodes are distributed in the 100m×100m and 200m×200m square field randomly.
- There areonly one base station and its located at the center of the field.
- Every sensor nodes and the base station are fixed after the distribution phase.
- Sensor nodes are capable of altering the transmission power according to the distance of the receiver nodes.
- All sensor nodes have equal energy when they are initially distributed.
- Each cluster has primary and backup cluster head nodes.
- For each collector node in the network, there is only one primary/backup cluster head to reach the BS.

In this paper, the energy consumption model proposed in [16, 17, 18] has been used.

## 3.1. An overview of IHR protocol

IHR is a distributed fault management technique, it considers two important aspects for a wireless sensor network, the first one is energy consumption and the second one is reliability. In this protocol, each collector node within a cluster has two cluster heads associated with it (PCH and BCH). Furthermore, each collector node only sent a data packet to the backup one when it discovered that the main cluster head failed, instead of transferring a data packet to the main cluster head and backup cluster head simultaneity. Therefore, the data transmitting process will not be interrupted during the lifetime of the network. Figure 1 illustrates the network model for the IHR protocol.

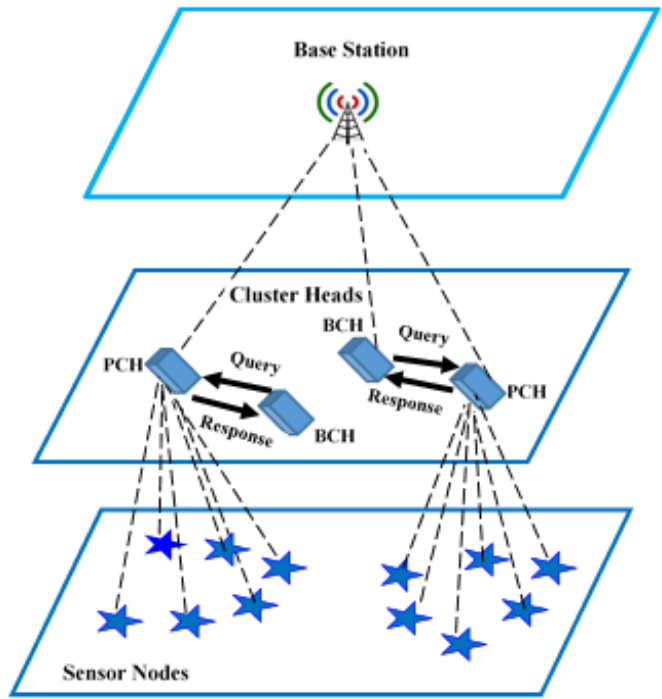

Figure 1. The network model for IHR protocol

IHR protocol considers hardware faults, particularly main cluster head failure. The hardware faults may be caused by receiver/transmitter failure, battery exhaustion, malicious attack, and malicious human activities. In this protocol, the impact of hardware failure is considered, because when amain cluster head gets fail, its relevant sensor nodes are disconnected. This will meaningfully decrease the availability and reliability of the sensor network.

## 3.2. An overview of the DHR protocol

DHR protocol considers an important aspect of wireless sensor network which is reliability. In this protocol, each collector node within a cluster has two cluster heads associated with it (PCH and BCH). Furthermore, each collector node sends a data packet to both of them at the same time. Consequently, aggregated data packets are forwarded to the base station through primary cluster heads as well as backup cluster heads. This protocol can guarantee data loss issue while a primary cluster head gets fail. A major drawback of DHR protocol is a duplication of forwarding every data packet over two disjoint paths towards the base station. Therefore, in data packet transferring procedure, energy will consume more and the overall network lifetime will be decreased.

## 4. PROPOSED FIHR PROTOCOL

The fuzzy IHR protocol is implemented based on the IHR protocol. The difference between IHR and FIHR is that the IHR protocol goes through the probabilistic model for clustering but FIHR protocol goes through the proposed fuzzy unequal clustering scheme. Furthermore, the distance between cluster heads has been considered to provide clustering in a distributed manner. The following pseudocode represented how the FIHR algorithm is implemented.

**Pseudocode**: Fuzzy Informer Homed Routing

**Input:** Network setting configuration
        N ← Number of deployed sensor nodes
        R ← Number of Rounds
        M← Number of times for data packet transmission

**Output:** FIHR Scheme
1:    T← probability to become a tentative CH
2:    Node Status← Non Cluster Head (NCH)
3:    clusterMembers← Null
4:    myClusterHead← Null
5:    beTentativeCH← True
6:    inquiry counter ← Null
7: for rounds.index=1:1: R do
8: x ← rand (0, 1)
9:    if x < T then
10: Measure distance to BS for Candidate CH
11: Calculate ComR of Candidate CH applying fuzzy logic (Input: Distance, Energy)
12: if ComR> Threshold
13:    beTentativeCH ← False
14:    else
15:      Calculate distance to other PCHs
16:      if ComR<=distance-ComR
17:        Node Status ← Primary Cluster Head (PCH)
18:        Advertise PCHmessage (ID, ComR) to other nodes
19:      else
20:        beTentativeCH ← False
21:      end if
22:    end if
23:      On receiving all PCHmessages
24:      myClusterHead ← The nearest PCH
25:      Send JoinMessage(ID) to the closest PCH
26:      Each PCH will select the node with the higher energy left among its entire cluster Members as the BCH.
27:      After selection of the BCHs, each PCH will inform its cluster members about BCH.
28: end if
29:    for times index =1:1: M do
30:      Each BCH starts inquiry from its relevant PCH for checking the aliveness.
31:      inquiry counter ← inquiry counter+1.
32:    if PCH is still working then
33:      PCH receives inquiry message from BCH then    responds to confirm its aliveness.
34:      Upon receiving the respond message from PCH, inquiry counter ← inquiry counter-1
35:     end if
36:    if inquiries counter > 3 then
37:      BCH decides the relevant PCH has failed then
38:      it sends inform message to its NCHs that send data packet to itself afterwards.
39:     end if
40:    if NCH receives the informer message from BCH then
41:      dispatches data packet to its relevant BCH.
42:     else
43:      dispatches data packet to its relevant PCH.
44:    end if
45:   end for
46: end for

In this protocol, there are two fuzzy input variables; the first one is residual energy and the second one is the distance to thebase station. Furthermore, the communication range (ComR) stands for the fuzzy output variable. If a sensor node has selected tocontribute in the competition, it becomes a tentative CH then in the local area competes to become primary cluster head. Furthermore, the selection of a primary cluster head is not only based on fuzzy input variables. The distance between primary cluster heads is also considered to increase the performance of the network. Moreover, each primary cluster head will select a sensor node with the highest energy left from its cluster members as a backup. Each primary cluster head will be check for aliveness by its backup cluster head. Figure 2 shows the schematic chart of proposed FIHR protocol.

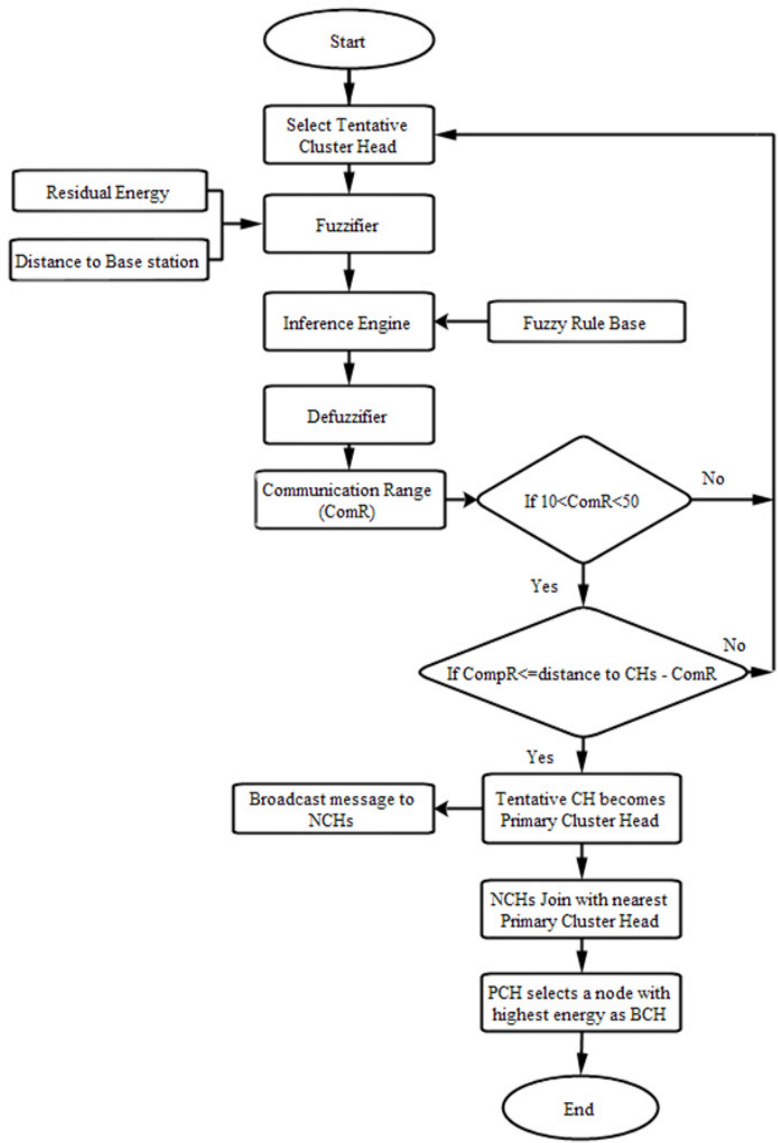

Figure 2. Schematic chart of the proposed FIHR protocol

## 5. CLUSTERING WITH FUZZY LOGIC SYSTEM

In this section, the proposed clustering piece of FIHR protocol using fuzzy logic is described in details. It is a distributed unequal clustering. It makes the local decision for determining communication range which leads to selecting primary cluster heads. Furthermore, it takes benefits of the fuzzy logic scheme to calculate the communication range of candidate primary cluster heads.

In the FIHR protocol, selecting a tentative primary cluster head is based on a probabilistic scheme. To estimate the communication range for a tentative primary cluster head, it employs both distances to BS and residual energy factors.
The communication range of each candidate primary cluster head is determined by using predefined fuzzy if-then rules to handle the uncertainty. Furthermore, to evaluate the rules, the Mamdani method, which is one of the most frequently used methods, is used as a fuzzy inference technique. The Center of Area (COA) scheme is employed for defuzzificationof the communication range [4].

The fuzzy logic system contains four modules [18, 10, 19]; fuzzification, fuzzy inference system, fuzzy rule base, and defuzzification with input and output variables. In the introduced method there are two crisp values as inputs (distance to BS and residual energy) and one output crisp value (Communication range). The basic model for fuzzy logic system incorporated in the proposed FIHR protocol is shown in Figure 3.

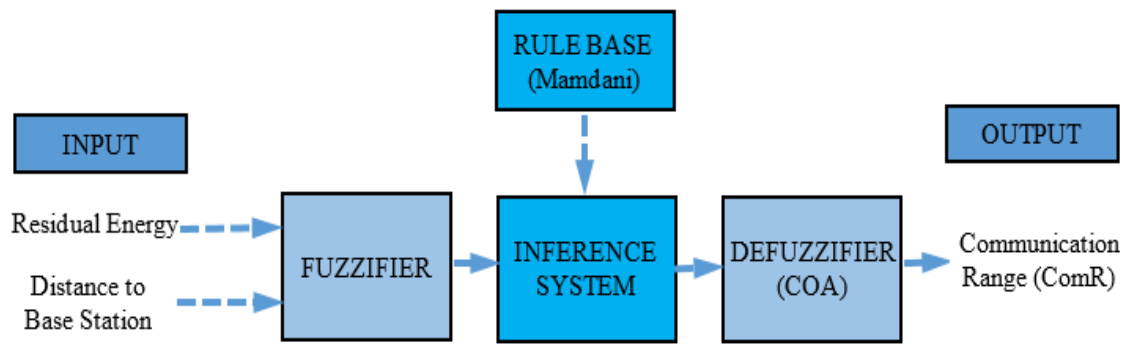

Figure 3. The basic model of the fuzzy logic system in FIHR protocol

- Fuzzification:Itstands for converting crisp input values into fuzzy sets through membership functions.
- Fuzzy rule base: It is used for storing If-Then rules.
- Fuzzy inference engine: It is used for combining fuzzy input values and fuzzy rules to simulate the reasoning by which it produces a fuzzy inference.
- Defuzzification:It stands for converting fuzzy outputs into crisp values.

For simplicity, trapezoidal and triangular membership functions are used in the proposed FIHR protocol. The mathematical formula for triangular membership function is described as in equation (1) [24].

$$\text{Triangular}(m; x, z, y) = \begin{cases} 0 & if & m < x \\ \frac{m-x}{z-x} & if & x \leq m \leq z \\ \frac{y-m}{y-z} & if & z \leq m \leq y \\ 0 & if & y \leq m \end{cases} \quad (1)$$

The parameters $\{x, z, y\}$ while $(x < z < c)$, determine the $m$ coordinates of the three angles of the underlying triangular membership function.

There are two types of a trapezoidal function, which are named leftmost trapezoidal and rightmost trapezoidal functions. These are defined in terms of coordinates $\{a, b, c, d\}$, where $a$ is upper support limit, $b$ is the upper limit, $c$ is lower limit and $d$ is lower support limit. The mathematical formula for leftmost and rightmost trapezoidal membership functions are described as in equations (1) and (2).

$$\text{Rightmost trapezoidal}(m; a, b) = \begin{cases} 0 & if & m > b \\ \frac{b-m}{b-a} & if & a \leq m \leq b \\ 1 & if & m < a \end{cases} \quad (2)$$

In case of Rightmost trapezoidal, the coordinate $a$ is less than coordinate $b$, while $c$ and $d$ are $-\infty$.

$$\text{Leftmost trapezoidal}(m; c, d) = \begin{cases} 0 & if & m > c \\ \frac{m-c}{b-a} & if & c \leq m \leq d \\ 1 & if & m < d \end{cases} \quad (3)$$

In case Leftmost trapezoidal the coordinate $c$ is less than coordinate $d$, while $a$ and $b$ are $+\infty$.

Figure 4 depicts an example of the triangular membership function.

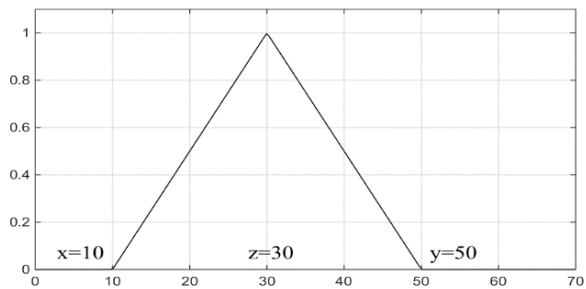

Figure 4. Triangular membership function

Figure 5 represents an example of the rightmost trapezoidal membership function and Figure 6 illustrates leftmost trapezoidal membership function of the proposed FIHR protocol.

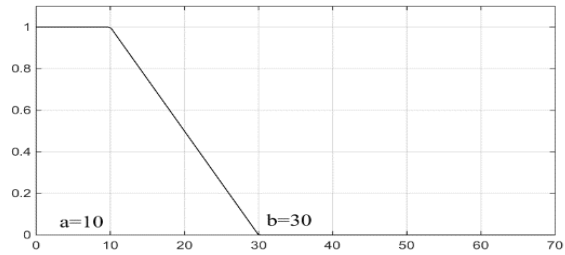

Figure 5. Rightmost trapezoidal membership function

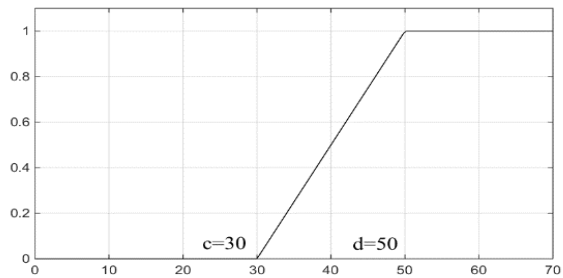

Figure 6. Leftmost trapezoidal membership function

The fuzzy set that defines the distance to thebase station as an input variable is represented in Figure 7. The linguistic variables for distance to BS are *close*, *med* and *far*. A trapezoidal membership function is preferred for *close* and *far* and a triangular membership function is preferred for*med*.

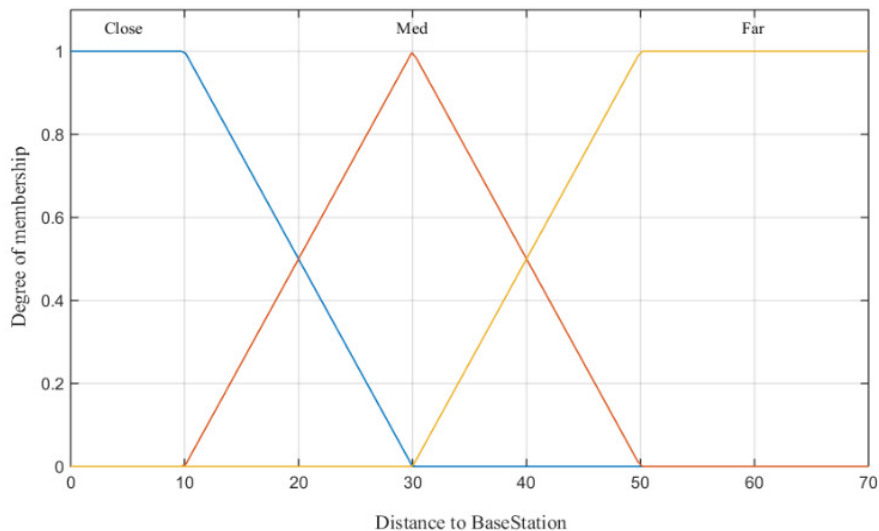

Figure 7. Membership functions of distance to base station

Moreover, the fuzzy set that defines the residual energy as another input variable is demonstrated in Figure 8. The linguistic variables for this fuzzy set are *low*, *med* and *high*. The *low* and *high* linguistic variables have a trapezoidal membership function while *med* linguistic variable has a triangular membership function.

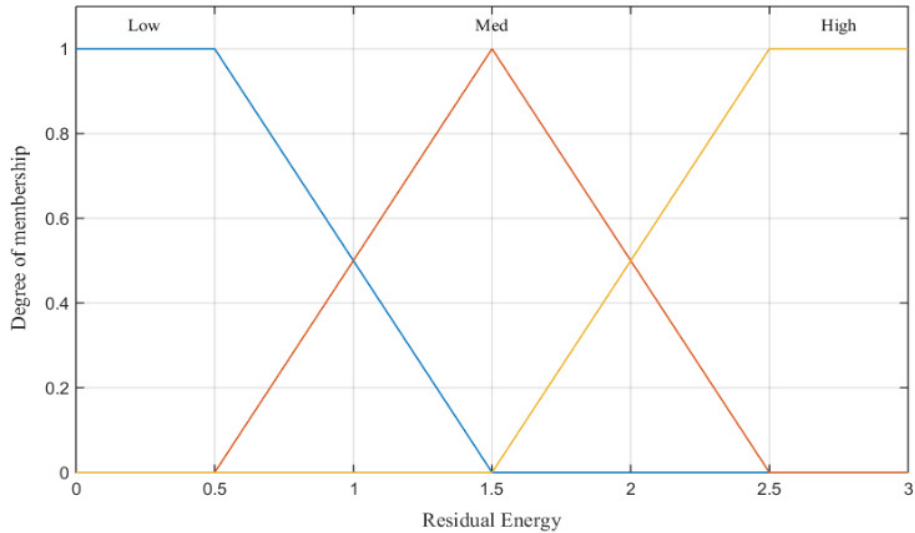

Figure 8. Membership functions of residual energy

The fuzzy output variable is the communication range of the tentative primary cluster head. The fuzzy set for the communication range is represented in Figure 9. The linguistic variables for this fuzzy set are *very small*, *small*, *rather small*, *med-small*, *med*, *med-large*, *ratherlarge,large* and *very large*. A trapezoidal membership function is preferred for *very small* and *very large* variables. The remaining linguistic variables are represented by using the triangular membership function.

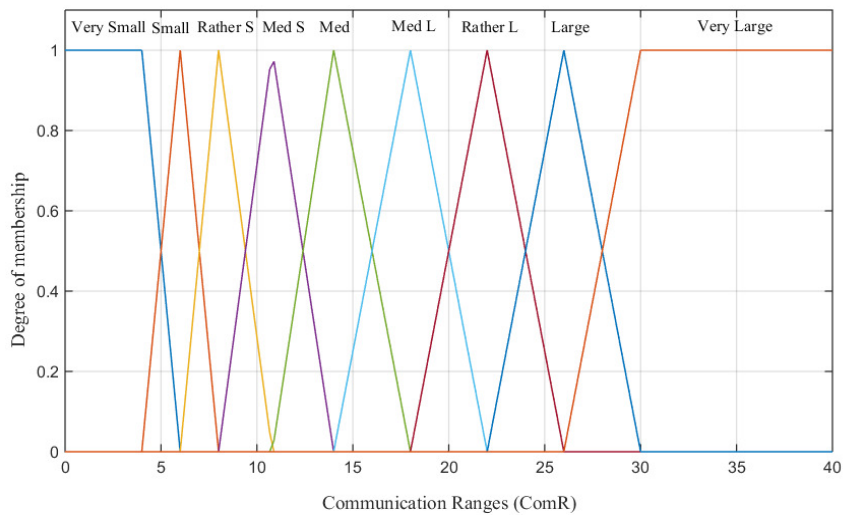

Figure 9. Membership functions of communication range

One of the main modules of the fuzzy inference system is the if-then rules which are designed to simulate the real world behavior [18]. The rules are typically based on knowledge of experts and experience on the same domain [8]. After the fuzzification step, the obtained membership values are applied to if-then rules for determining fuzzy output set. Table 1 represents the possible rules that are considered in the proposed FIHR protocol for calculating the communication range of a

52

tentative PCH. If a selected tentative PCH located far away from the base station and its energy is full, then it has the highest communication range (very large). On the other hand, if its energy is very low and it is the nearest sensor node to the base station, then it has the lowest communication range (very small). Furthermore, the remaining intermediate possibilities fall among these two extreme cases.

Table 1. Possible if-then rules

| Rule No. | Fuzzy input variables | | Fuzzy output variable |
|---|---|---|---|
| | Residual energy (J) | Distance to BS (m) | ComR |
| 1 | Low | Close | Verysmall |
| 2 | Med | Close | Small |
| 3 | High | Close | Rathersmall |
| 4 | Low | Med | Medsmall |
| 5 | Med | Med | Med |
| 6 | High | Med | Medlarge |
| 7 | Low | Far | Ratherlarge |
| 8 | Med | Far | Large |
| 9 | High | Far | Verylarge |

In the defuzzification step, the Center of Area (COA) scheme is employed for a crisp output value (ComR). The relationship between input variables (residual energy and distance to base station) and the output variable (communication range) is represented in Figure 10.

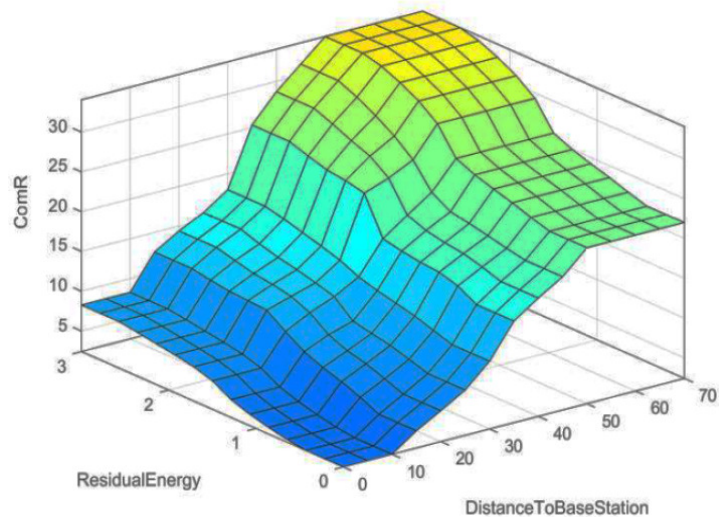

Figure 10. The relationship between input and output variables in the proposed FIHR fuzzy logic system

## 6. SIMULATION RESULTS AND DISCUSSIONS

To evaluate the performance of the proposed protocol, two different scenarios are implemented. In the first scenario, hundred sensor nodes are randomly spread in the area size of (100×100) $m^2$ anda base station is placed at the center of the network, at (50, 50) coordinates. In the second

53

scenario, two hundred sensor nodes are randomly spread in the area size of (200×200) m$^2$ and a base station is placed at the center of the network, at (100, 100) coordinates. The simulation parameters are summarized in Table 2.

Table 2. Simulation parameters

| Parameter | Value |
|---|---|
| Network size (meters) | (100m×100m), (200m×200m) |
| Location of BS | (50, 50), (100, 100) |
| Number of Sensor nodes | 100, 200 |
| Sensor Nodes Energy | 3.0 J |
| ETx ($E_{elec}$) / ERx ($E_{elec}$) | 50 nJ/bit |
| Data packet size | 32000 bits |
| Number of data transmission per round | 3 |
| Query/Respond messages size | 160 bits |
| $\varepsilon_{fs}$ and $\varepsilon_{mp}$ | 10, 0.004   pJ/bit/m$^2$ |
| Eda ($E_{Aggregation}$) | 5    nJ/bit/signal |

The numberof researchers used the metrics First Node Dead (FND) and Half of Nodes Alive (HNA) to evaluate the performance of the network [4, 20, 21, 22] and to estimate the lifetime of the network [4]. The metric FND denotes an estimated value for the round in which the first sensor node dies. Furthermore, the metric HNA denotes an estimated value for the round in which the half of sensor nodes dies [23]. Therefore, the metrics FND and HNA isused to evaluate the effectiveness of the proposed protocol over the IHR and DHR protocols. To yield more reliable outcomes, the experiments were conducted 20 times and the averages of the results have been taken for each protocol. Furthermore, to assess the performance of the offered protocol over the IHR and DHR protocols, two different scenarios are developed.

### 6.1. Scenario 1

In this scenario, hundred sensor nodes are randomly spread in the area size of (100×100) m$^2$ and the base station is placed at the center of the network at (50, 50) coordinates. The detailed configuration of this scenario is presented in Table 2. The simulation of the proposed method, IHR and DHR schemes yielded the following results:

Table 3 shows the measured values of FND and HNA metrics concerning the number of rounds till which the first sensor node gets dead and half of the sensor nodes are alive for simulated protocols.

Table 3. Values of FND and HNA metrics

| Protocol | FND | HNA |
|---|---|---|
| DHR | 106 | 166 |
| IHR | 123 | 289 |
| FIHR | 126 | 304 |

From Table 3 can be observed, the proposed method (FHIR) outperforms IHR and DHR schemes in both FND and HNA metrics. As to evaluate the energy efficiency of the proposed method, the total residual energy metric is used. The total residual energy of the network is calculated after each round. Table 4 shows the total residual energy levels for the proposed method, IHR and DHR schemes at various rounds. Since each sensor node has 3J initial energy level, the entire energy level of the network is equal to 300J in scenario 1 at the beginning simulation time.

Table 4. Total residual energy levels for DHR, IHR and FIHR protocols

| Round No. Protocol | 50 | 100 | 200 | 300 | 400 |
|---|---|---|---|---|---|
| DHR | 206.56 | 113.11 | 0 | 0 | 0 |
| IHR | 249.48 | 197.78 | 100.86 | 38.37 | 2.03 |
| FIHR | 250.89 | 200.61 | 106.24 | 40.04 | 4.95 |

From the above table, it can be concluded that the FIHR method can save a little bit more energy as compared with the IHR method and significantly save energy as compare with DHR scheme in this scenario. Table 5 represents the throughput value during the lifetime of the network for simulated protocols. From the table, it can be concluded that the FIHR protocol can deliver packets a little bit more as compared with IHR and 22170 KB data packets more than DHR protocol in scenario 1.

Table 5. Throughput for FIHR, IHR and DHR protocols

| Protocol | Packet delivery (KB) |
|---|---|
| DHR | 19590 KB |
| IHR | 41625 KB |
| FIHR | 41760 KB |

The parameter, number of dead nodes has also been considered for performance evaluation of the network. Figure 11 represented the number of dead nodes for FIHR, DHR and IHR protocols at various rounds. From the results, it can be observed that the proposed FIHR protocol outperforms IHR and DHR protocols.

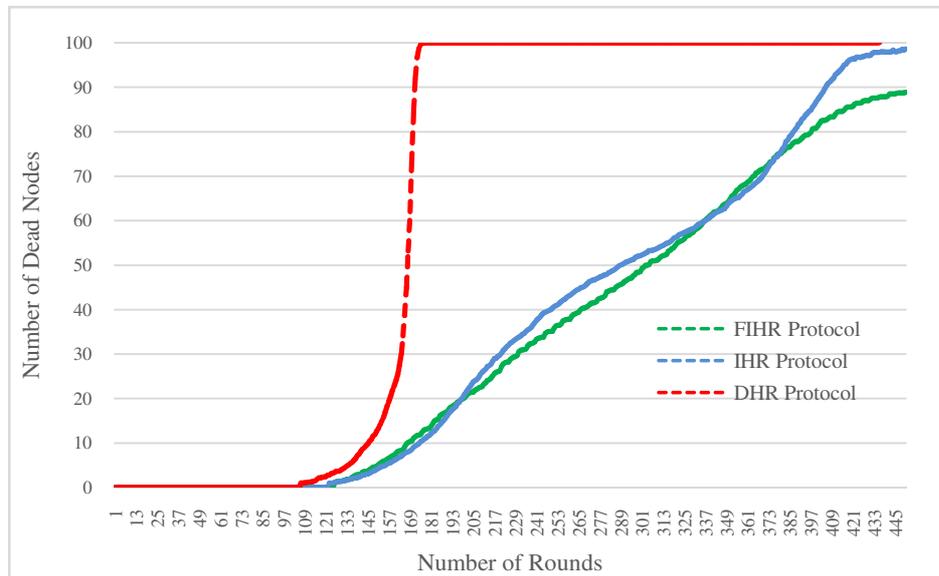

Figure 11. Number of dead nodes for FIHR, IHR and DHR protocols at various rounds

## 6.2. Scenario 2

In this scenario, two hundred sensor nodes are randomly spread in the area size of (200×200)m$^2$ and the base station is placed at the center of the network at (100, 100) coordinates. The detailed configuration of this scenario is presented in Table 2. The simulation of the proposed method, IHR and DHR protocols yielded the following results.

As seen in Table 6 the proposed FIHR protocol performs better than IHR and DHR protocols for both FND and HNA metrics. FIHR protocol is 42.6% and 123.33% more efficient than IHR and DHR protocols respectively in terms of FND. Moreover, it is 29.7% and 127.66% more efficient than IHR and DHR protocols while the HNA metric is considered for performance evaluation.

Table 6. Values of FND and HNA metrics

| Protocol | FND | HNA |
|---|---|---|
| DHR | 30 | 94 |
| IHR | 47 | 165 |
| FIHR | 67 | 214 |

Figure 12 demonstrates the number of dead nodes concerning the number of rounds for both protocols. This figure displays that the FIHR protocol is steadier than the IHR and DHR protocols. Because the workload is distributed among cluster heads and sensor node deaths begin later for FIHR protocol and continue linearly till all sensor nodes die in the network.

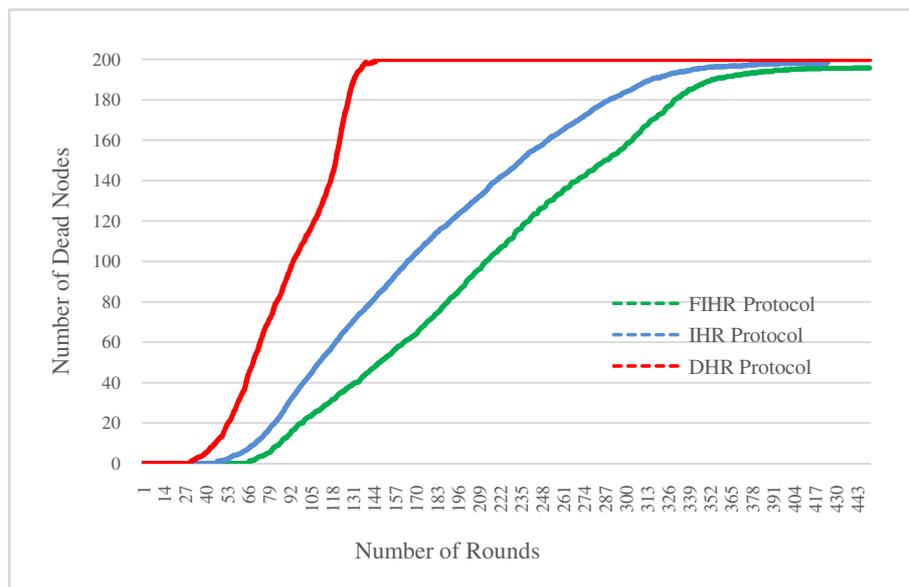

Figure 12. Number of dead nodes for FIHR, IHR and DHR protocols at various rounds

As to evaluate energy the efficiency of the FIHR protocol, the total remaining energy metric is used. Figure 13 shows the total remaining energy levels for proposed FIHR, DHR and IHR protocols at various rounds for scenario 2. Since each sensor node has 3J initial energy and the number of sensor nodes is200, the total energy level of the network is equal to 600J at the beginning simulation time.

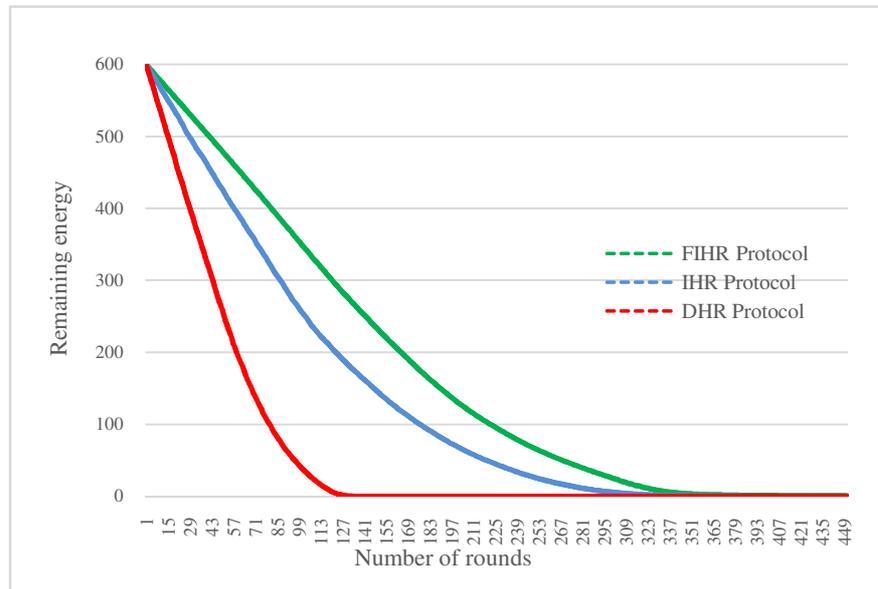

Figure 13. The total remaining energy levels for DHR, IHR and FIHR protocols at various rounds

From the above figure, we can conclude that the FIHR protocol can save more energy as compared to IHR and DHR protocols in this scenario.

Table 7 represents the throughput value during the lifetime of the network for DHR, IHR and FIHR protocols. From the table, it can be concluded that the FIHR protocol can deliver packets more as compared with IHR and DHR protocols. Moreover, it is 9.14% and 158.63% more efficient than IHR and DHR protocols respectively in scenario 2.

Table 7. Throughput for FIHR and IHR protocols

| Protocol | Packet delivery (KB) |
|---|---|
| DHR | 14451 KB |
| IHR | 34246 KB |
| FIHR | 37375 KB |

## 7. CONCLUSION

The paper presented a distributed fault tolerance mechanism for wireless sensor networks. The method designed and developed based on IHR protocol. It is called Fuzzy IHR fault tolerance protocol. The difference between IHR and FIHR is that the IHR protocol goes through the probabilistic model for clustering but FIHR protocol goes through the proposed fuzzy unequal clustering scheme. Fuzzy IHR protocol tries to distribute the workload between every sensor node. To achieve this objective, it assigns appropriate communication range to the selected tentative primary cluster head using fuzzy logic. The fuzzy parameters to calculate communication range values of tentative primary cluster heads are remaining energy and distance tothe base station. The parameter distance between cluster heads also is considered to avoid overlapping of them and obtaining distributed workload of cluster heads. According to the simulation results the FIHR has better performance compared to IHR and DHR protocols in terms of first node dead, half of the nodes alive, throughput and total energy remaining of the network. Furthermore, results imply that the workload is distributed and the sensor nodes tend to fail later within the lifetime of the network. Moreover, the remaining energy level of the network in FIHR protocol at a certain round is higher than with IHR and DHR protocols. As a result, FIHR protocol is more energy efficient than the IHR and DHR protocols. It is concluded that the proposed protocol is a stable and energy efficient fault tolerance algorithm for WSNs.

As FIHR protocol is introduced for WSNs that containing stationary sensor nodes, it can be extended for handling mobile sensor nodes. Furthermore, it can be extended for large scale networks while using two or more levels of clustering.

## REFERENCES

[1] Atero,F. J., Vinagre,J. J.,Ramiro J. and Wilby,M. "A low energy and adaptive routing architecture for efficient field monitoring in heterogeneous wireless sensor networks," In proc. IEEE Int. Conf. on High Performance Computing and Simulation, Istanbul, Turkey, 2011, pp.449 – 455.

[2] Baroutis, N. and Younis, M., (2017)"Load-conscious maximization of base-station location privacy in wireless sensor networks," Computer Networks, 124, pp.126-139.

[3] Kakamanshadi, G., Gupta, S. and Singh, S. "A survey on fault tolerance techniques in Wireless Sensor Networks," In proc. IEEE Int. Conf. on Green Computing and Internet of Things, India, 2015, pp. 168-173.

[4] Biswas, S., Das, R. and Chatterjee, P., (2018)"Energy-efficient connected target coverage in multi-hop wireless sensor networks," In Industry Interactive Innovations in Science, Engineering and Technology, Springer, Singapore, pp. 411-421.

[5] Bagci, H. and Yazici, A., (2013)"An energy aware fuzzy approach to unequal clustering in wireless sensor networks," Applied Soft Computing, vol. 13, no.4, pp.1741-1749.

[6] Mittal, N., Singh, U. and Sohi, B.S., (2017)"A stable energy efficient clustering protocol for wireless sensor networks,"Wireless Networks, vol.23, no.6, pp.1809-1821.

[7] Pachlor, R. and Shrimankar, D., (2018)"EEHCCP: an energy-efficient hybrid clustering communication protocol for wireless sensor network," In Ad Hoc Networks, Springer, Cham, vol.223, pp. 199-207.

[8] Wang, J., Niu, J., Wang, K. and Liu, W., (2018)"An energy efficient fuzzy cluster head selection algorithm for WSNs," In IEEE International Workshop on Advanced Image Technology (IWAIT), pp. 1-4.

[9] Tamandani, Y.K., Bokhari, M.U. and Shallal, Q.M., (2017)"Two-step fuzzy logic system to achieve energy efficiency and prolonging the lifetime of WSNs," Wireless Networks, vol.23, no.6, pp.1889-1899.

[10] Torghabeh, N. A., Totonchi, M. R. A. and Moghaddam, M. H. Y. "Cluster Head Selection using a Two-Level Fuzzy Logic in Wireless Sensor Networks," In proc. 2nd IEEE Int. Conf. on Computer Engineering and Technology, China, 2010, pp.357–361.

[11] Messaoudi, A., Elkamel, R., Helali, A. and Bouallegue, R.,"Distributed fuzzy logic based routing protocol for wireless sensor networks," In proc. 24thIEEE Int. Conf. onSoftware, Telecommunications and Computer Networks (SoftCOM), 2016,pp. 1-7.

[12] Arjunan, S. and Sujatha P., (2017)"A survey on unequal clustering protocols in Wireless Sensor Networks,"Journal of King Saud University-Computer and Information Sciences,vol. xx, PP.1-14.

[13] Meikang, Q., Ming, Z., Li, J., Liu, J., Quan, G. and Zhu, Y., (2013)"Informer homed routing fault tolerance mechanism for wireless sensor networks," Journal of Systems Architecture, vol. 59, pp.260-270, May. 2013.

[14] Jain, N., Vokkarane, V.M. and Wang, J.P., "Performance analysis of dual-homed fault tolerant routing in wireless sensor networks," In proc. IEEE Conf. on Technologies for Homeland Security, Waltham, 2008, pp. 474–479.

[15] Abedi, R. H., Aslam, N. and Ghani, S., "Fault tolerance analysis of heterogeneous wireless sensor network," In proc. 24th IEEECanadian Conf. on Electrical and Computer Engineering, Canada, 2011, pp.175-179.

[16] Akyildiz, F. and Vuran, C. Wireless sensor networks. 1st ed. Georgia, Institute of Technology, USA: John Wiley & Sons, 2010, pp.48-49.

[17] Xiaoling, W., Cho, J., Brian, J. d. and Lee, S., "Energy-aware routing for wireless sensor networks by AHP," In IFIP Int. Workshop on Software Technologies for Embedded and Ubiquitous Systems, Springer Berlin Heidelberg, 2007, pp. 446-455.

[18] Masdari, M. and Naghiloo, F., (2017)"Fuzzy Logic-Based Sink Selection and Load Balancing in Multi-Sink Wireless Sensor Networks," Wireless Personal Communications, vol.97, no.2, pp.2713-2739.

[19] El Alami, H. and Najid, A., "March. Energy-efficient fuzzy logic cluster head selection in wireless sensor networks," In proc.IEEE Int. Conf. on Information Technology for Organizations Development (IT4OD), 2016, pp. 1-7.

[20] Agrawal, D. and Pandey, S., (2017)"FUCA: Fuzzy-based unequal clustering algorithm to prolong the lifetime of wireless sensor networks," International Journal of Communication Systems, vol.31, no.2, pp.1-18

[21] Shankar, T., Karthikeyan, A., Sivasankar, P., and Rajesh A., (2017)"Hybrid approach for optimal cluster head selection in WSN using leach and monkey search algorithms," Journal of Engineering Science and Technology, vol.12, no.2, pp.506-517.

[22] Adhikary, D.R.D. and Mallick, D.K., (2017)"An Energy Aware Unequal Clustering Algorithm using Fuzzy Logic for Wireless Sensor Networks," Journal of ICT Research and Applications, vol.11, no.1, pp.55-76.

[23] Sert, S.A., Bagci, H. and Yazici, A., (2015)"MOFCA: Multi-objective fuzzy clustering algorithm for wireless sensor networks," Applied Soft Computing, vol.30, pp.151-165.

[24] Logambigai, R. and Kannan, A., (2016)"Fuzzy logic based unequal clustering for wireless sensor networks,"Wireless Networks, vol.22, no.3, pp.945-957.

## AUTHORS BIOGRAPHY

**GholamrezaKakamanshadi** received the M.Tech. degree in Computer Engineering from BharatiVidyapeeth University, Pune, India, in 2012. He is currently pursuing his Ph.D. in the Department of CSE, University Institute of Engineering & Technology, Panjab University, Chandigarh, India. His research interests are Wireless sensor network, Body area network, Internet of Things and Network security.

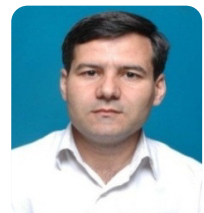

**Savita Gupta** received the B.Tech and M.E. degree in both Computer Science and Engineering. She obtained her Ph.D. degree in 2007 in the field of Medical Image Processing. She has been into the teaching profession since 1992 and has published more than 70 papers in refereed International Journals and conference proceedings. Presently, she is working as Director in the University Institute of Engineering & Technology, Panjab University, Chandigarh, India. She has completed various research projects funded by various agencies like DST, AICTE, and MHRD. Her research interests include Medical image processing, Wavelet based image compression and denoising, Network security and Wireless sensor networks.

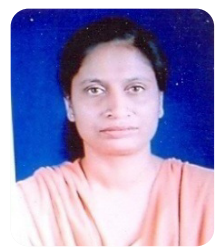

**Sukhwinder Singh** received the B.Tech degree in 1991 and M.E. degree in 1999. he obtained his Ph.D. degree in 2006 from IIT, Roorkee. Presently, he is working as a professor in the Department of CSE, University Institute of Engineering & Technology, Panjab University, Chandigarh, India. His research interests include Medical Image Processing and Analysis, Biomedical Signal Processing, Information Retrieval, Wireless Sensor Networks, Body Area Networks, Cognitive Enhancement, Machine Intelligence.

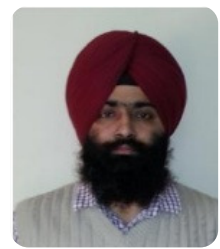